\documentstyle[12pt]{article}

\newcommand{\be}{\begin{equation}}
\newcommand{\ee}{\end{equation}}
\newcommand{\ba}{\begin{eqnarray}}
\newcommand{\ea}{\end{eqnarray}}
\newcommand{\fisfa}{\Phi_{sph}}

\newcommand{\omeg}{\Omega}
\begin{document}
\begin{flushright}
{PITT-93-6}
\end{flushright}
\begin{center}
{\bf REAL TIME ANALYSIS OF THERMAL ACTIVATION}\\
{\bf VIA SPHALERON TRANSITIONS}
\end{center}
\begin{center}
{{\bf D. Boyanovsky$^{(a)}$ and C. Arag\~{a}o de Carvalho$^{(b)}$}}
\end{center}
\begin{center}
{\it (a) Department of Physics and Astronomy\\ University of
Pittsburgh, Pittsburgh, P. A. 15260, U.S.A.} \\
{\it (b) Departamento de Fisica, Pontificia Universidade Catolica\\
C.P. 38071, CEP 224, Rio de Janeiro, Brazil.}
\end{center}
\begin{abstract}
We study the process of thermal activation mediated by sphaleron
transitions by analyzing the real-time dynamics of the decay out of
equilibrium in a $1+1$ dimensional field theory with a metastable state.
The situation considered is that of a rapid supercooling  in which the
system is trapped in a metastable state at a temperature larger than the
mass of the quanta, but smaller than the energy
to create a critical droplet. The initial density matrix is evolved in
time and the nucleation rate (probability current at the saddle point)
 is computed.
The nucleation rate is {\it time dependent}, vanishing at early times,
 reaching
a maximum at a time $t \approx 1/m$ with $m$ the mass of quanta in the
metastable state,  and decreasing at long times as a
 consequence of unitarity.
An estimate for the average number of particles of ``true vacuum''
produced as
a function of time during the nucleation process is obtained.
\end{abstract}

\newpage

\section{\bf Introduction and Motivation}

Thermal activation plays a very important role in the dynamics of
evolution out of equilibrium of thermodynamically metastable states.
 Such a situation arises for example in first order phase transitions
 which may have ocurred in the early
 universe\cite{linde1,stein,kolb,brandenberger},
where false vacuum decay provides a mechanism for ending inflation.
 More recently thermal activation has been
 conjectured
to be the mechanism responsible for unsuppressed
 baryon number violation at high
temperatures\cite{kuzmin,susskind,manton}.
Within the context of baryon number violation, Kuzmin, Rubakov and
Shaposnikov\cite{kuzmin}, argued that in
non-Abelian gauge theories topologically different vacua (characterized
by different Chern-Simons number) are separated by energy barriers
 and that there are configurations (in functional
space) that are responsible for {\it over the barrier}
 transitions from one vacuum to another.
Klinkhamer and Manton\cite{manton} suggested that these
 over-the-barrier transitions correspond to static configurations that
extremize the energy functional with at least one unstable mode. These
authors named this configuration the ``sphaleron'' and provided
a topological proof for the existence of such configurations in
 the Standard Model.
In non-Abelian gauge theories this configuration carries {\it half a unit}
of winding number, and corresponds to a {\it saddle point} in the space
of functions.  These transitions are necessarily
accompanied by a change in baryon
 number\cite{mottolawipf,mclerran,dineetal,dinesakita}.

Within the context of first order phase transitions at finite temperature,
Linde\cite{linde2} argued that the relevant configuration responsible for
the decay of the metastable state
 is an $O(3)$
(static) ``bounce'', which is in a sense the high temperature limit of the
zero temperature bounce that describes a tunneling event\cite{coleman}
between the metastable and stable phases.

An important quantity is the {\it transition rate} (per unit volume),
as this quantity determines the relevant time scales for the completion
of first order phase transitions in inflationary cosmologies and for the
rate of baryon number violation in gauge theories at finite temperature.

The usual calculation of the rate follows Langer's original theory of
homogeneous nucleation in classical statistical mechanics\cite{langer},
 and  Affleck's\cite{affleck} generalization to the quantum case.

Langer's  treatment is based on the computation  of the
probability current at the saddle point corresponding
to a {\it steady state} solution of an appropriate Fokker-Planck equation.
It is assumed in this treatment, and clearly spelled out in Langer's
calculation, that the steady state condition corresponds to the situation
in which the metastable phase is {\it replenished} at the same rate
at which probability is flowing out across the saddle. There is a
``source'' of probability on the metastable well and a corresponding
``sink'' on the stable side. The saddle point configuration corresponds
to a ``critical droplet''. The probability flows from the metastable to
the stable phase along the unstable direction in functional space.
 Under the steady-state assumption the rate is calculated in
{\it equilibrium} and identified to be proportional to the imaginary
part of an {\it analytically continued} free energy.
The same assumptions are implicit in Affleck's approach in which the
transition rate is computed as an equilibrium average\cite{affleck}.
 As a result of the
steady-state assumption, the rate is independent of time and
basically determined by an Arrhenius (thermal activation) equilibrium
 Boltzmann factor\cite{hanggi}.

Langer's treatment of nucleation has been recently generalized to
 relativistic field theory (under basically the same assumption of a steady
state) by Csernai and Kapusta\cite{kapusta}.

 The ``critical droplets''
in Langer's theory of homogeneous nucleation are thus identified with
the ``sphalerons'' in the broader sense studied by Manton and
 Samols\cite{manton2}, i.e,  static field configurations with (at
least) one unstable mode. These are saddle points of the energy
functional corresponding to field configurations ``sitting'' at the
top of the energy barrier (in functional space).
Numerical simulations of sphaleron transitions have been carried out
by Hellmund and Kripfganz\cite{krip} and Ambjorn et. al.\cite{ambj}.
 These authors integrate numerically
the {\it classical} equations of motion by starting from a sphaleron
 configuration.
Recently a numerical study of the process of thermal activation
assuming a simple Langevin description has been reported\cite{gleiser}.

The motivation of this article is to provide a {\it different}
approach to the study of the decay of a metastable state
by considering the process of
thermal activation through ``sphaleron transitions'' as a non
equibrium evolution in {\it real time}.

We consider
the situation of a $1+1$ dimensional scalar field theory in
 which a metastable state has been formed after a
period of rapid supercooling and decays via the formation of ``critical
droplets'' and their subsequent evolution.
We propose to describe the supercooled metastable state
in terms of a functional density matrix corresponding to a
thermal distribution of free-field quanta (harmonic oscillator states)
of the false vacuum. The temperature of the ensemble is assumed to be
much larger than the mass of the quanta ($m$) but smaller than the
energy to create a sphaleron configuration.
 The ``sphaleron'' determines an unstable direction
in functional space along which probability flows.
We solve for the {\it real time evolution}
of the initial density matrix along this unstable direction. The
transition rate is identified with the probability current flowing
across the saddle from the metastable to the stable state.  This
approach provides a description of the decay process {\it out of
equilibrium} from an initial state.
 We find that the transition rate is time dependent
and that it strongly depends on the initial state.

In section II we study the properties of the ``sphaleron''
and quantize the theory around this static semiclassical  configuration.
Section III presents a detailed analysis of the time evolution of the
field theoretical density matrix and the probability current along the
unstable direction on the saddle point is evaluated. Following
Langer's original treatment,  this probability current is identified
with the transition rate. In this section we also mention some recent
experiments on nucleation in classical fluids that report a time
 dependent rate, and dependence on initial conditions.

In section IV we present an approximate calculation for the production
of particles of the true vacuum during the decay process.
Section V summarizes our conclusions, establishes the range of validity
of our results, and explores potential implications.

 Two appendices contain relevant technical details.

\section{\bf The Sphaleron and Thermal Activation:}

We consider a scalar field theory in $1+1$ space-time
dimensions with Hamiltonian

\ba
H       & = & \int dx \left\{\frac{1}{2}\Pi^2(x)+
\left(\frac{d\Phi(x)}{dx}\right)^2+
 V(\Phi(x))\right\} \label{hamiltonian}\\
V(\Phi) & = & \lambda (\Phi-\Phi_{-})^2 \Phi (\Phi-\Phi^{*})
 \label{potential}
\ea
The potential is depicted in figure 1(a).
 The point $\Phi^*$
is parametrized by the mass $m$ of harmonic oscillations around the minimum
$\Phi_{-}$ as
\be
\Phi^{*} = \Phi_{-}\left[1-\frac{m^2}{2\lambda\Phi_{-}^2}\right]
\label{fistar}
\ee
In two space time dimensions, the scalar field is dimensionless and the
coupling constant has dimensions of $(mass)^2$.

The ``sphaleron'' is defined as a static field configuration that
 corresponds to an extremum of the energy functional with
at least one {\it unstable} direction in functional space.
In our case, a stationary solution of the equations of motion (extremum of
the energy functional) satisfies
\be
-\frac{d^2 \Phi(x)}{dx^2}+\frac{\partial V(\Phi)}{\partial\Phi} =
 0 \label{eqofmotion}
\ee
This equation resembles that of a particle moving in ``time''
 (labeled by x) down a potential $-V$. The solution that starts at
 $\Phi_{-}$ at
$x \rightarrow -\infty$ and returns to $\Phi_{-}$ as
 $x \rightarrow \infty$ is easily found to be\cite{aragao2}
\ba
\fisfa   & = & \Phi_{-}+ \frac{m}{2\sqrt{2\lambda}}\left[\tanh
[\frac{m}{2}(x-x_o)+s_o]-\tanh[\frac{m}{2}(x-x_o)-s_o] \right]
\label{sfaleron}\\
s_o       & = & \frac{1}{2}\cosh^{-1}\left[\frac{\epsilon +1}{\epsilon -1}
\right] \label{so} \\
\epsilon & = & \frac{m^2}{2\lambda\Phi_{-}^2} \label{epsilon}
\ea

\noindent where $x_o$ is an arbitrary integration constant and
 reflects the
 translational invariance of the equations of motion.
This solution corresponds to a kink-antikink
pair or  ``droplet'' with a ``radius'' $2 s_o / m $, and is similar to
a ``polaron solution'' found in quasi-one dimensional
 polymers\cite{aragao1}. This solution and its stability was also
studied by Baltar et. al.\cite{llambias}

 The dimensionless ratio $\epsilon$ is a measure of
the depth of the global minimum ($\Phi_{+}$). As
$\epsilon \rightarrow 1$ the minima of the potential become degenerate and
  $s_o \rightarrow \infty$. The situation
 $\epsilon \approx 1; \; \; s_o \gg 1$
  corresponds to a ``thin-wall'' droplet for
which the radius is much larger than the wall thickness or
 ``skin'' of the
one-dimensional droplet $\xi = 2/m$, as depicted in figure 2.
For a ``thin-wall'' droplet we find the energy of the sphaleron
configuration to be
\be
E_{sph} = \frac{m^3}{6\lambda} + {\cal{O}}(e^{-2s_o})
 \label{sfaleronenergy}
\ee
where we have shown that the corrections are exponentially small in the
``thin-wall'' limit.

It is convenient to expand the field and its canonical momentum
 around the sphaleron configuration as
\ba
                             \Phi(x) & = & \fisfa(x-x_o)+ \sum_{l}
 \phi^{[x_o]}_{l}f_{l}(x-x_o) \label{fil} \\
                              \Pi(x) & = & \sum_{l} \pi^{[x_o]}_{l}
f_{l}(x-x_o) \label{pil} \\
\left[ \pi^{[x_o]}_{l} \; , \;  \phi^{[x_o]}_{l'}\right]
                                     & = & -i \hbar \delta_{l,l'}
 \label{canonical}
\ea

\noindent with the $f_l(x-x_o)$ being the eigenfunctions of
 the operator of quadratic
fluctuations around the sphaleron configuration
\be
\left[-\frac{d^2}{dx^2}+V''(\fisfa(x-x_o))\right]f_l(x-x_o) = \omega_l^2
f_l(x-x_o) \label{fl}
\ee
chosen to be real and orthonormal (in a volume L).
We have explicitly written the dependence of the eigenfunctions on $x_o$
as a consequence of translational invariance, and indicated that the
operators $\phi_l, \pi_l$ depend {\it parametrically} on $x_o$. This will
become important later when we introduce collective coordinates to treat
translational invariance.

Translational invariance guarantees that there is a zero frequency mode
whose normalized eigenfunction is given by
\be
f_0 (x-x_o) = \frac{1}{\sqrt{E_{sph}}}\frac{d\fisfa(x-x_o)}{dx}
\label{zeromode}
\ee
This eigenfunction is parity odd and has one node, and in the thin-wall
limit it is identified as the antisymmetric linear combination of the
zero modes for kink and anti-kink. There is an orthogonal symmetric
combination that is parity even and nodeless corresponding to the
orthogonal linear combination, which is the eigenfunction with the lowest
eigenvalue. In the thin-wall limit we find it to be
\be
f_{-1} = \frac{m}{2\sqrt{E_{sph}}} \frac{d\fisfa(x-x_o;s_o)}{ds_o}
\label{unstablemode}
\ee
This mode has  a {\it negative eigenvalue} $\omega_{-1}^2 =
-\Omega^2$ ($\Omega >0 $), in the ``thin-wall'' limit
 this eigenvalue is
exponentially small, determined by the overlap integral of the
kink-antikink zero modes $\Omega^2 \approx {\cal{O}}(e^{-2s_o})$.
In this approximation, and based on the known results on bound-states
of the one kink case, we conclude that the rest of the spectrum is
positive definite.

The Hamiltonian in this basis, to be referred to as the
 ``sphaleron basis'', becomes

\ba
  H & = & E_{sph}+H_q+ H_{I} \label{hamsfal} \\
H_q & = &  \frac{\pi_0^2}{2}+\frac{\pi_{-1}^2}{2}-\frac{\Omega^2}
{2}\phi_{-1} +\frac{1}{2}\sum_{l\geq 1}\left[\pi_{l}^2 + \omega_{l}^2
\phi_{l}\right] \label{hamsfal2}
\ea

\noindent where $H_{I}$, has terms cubic and quartic in terms of the $\phi_l$
and
we have suppressed the upper index $[x_o]$ in both the $\pi_l \; ;
\phi_l$ to avoid cluttering of notation, but keeping in mind that these
operators depend parametrically on $(x_o)$.

The nature of the instability represented by the mode $l=-1$ is physically
clear. For small amplitudes of $\phi_0\; ; \; \phi_{-1}$ the field may be
written as
\ba
\Phi(x) & \approx & \fisfa[x-(x_o+\delta x_o); (s_o+\delta s_o)]
 +\sum_{l\geq 1}f_l(x-x_o)
\phi_l \label{shiftedfisfa} \\
\delta x_o
        &   =     & \frac{\phi_0}{\sqrt{E_{sph}}}
         \; \; ; \; \;  \delta s_o = \frac{m}{2\sqrt{E_{sph}}}\phi_{-1}
\ea
 Whereas $\phi_0$ represents a translation of the
 center of mass of the
sphaleron configuration, $\phi_{-1}$ represents an expansion or contraction
of the radius of the droplet. Thus we identify the collective coordinate
that describes the radius of the droplet (in this approximation) as
\be
s = s_o + \frac{m}{2\sqrt{E_{sph}}} \phi_{-1} \label{radio}
\ee
The coordinate $\phi_{-1}$ measures the departure from the critical
droplet (for small amplitudes of this coordinate).

 When the droplet expands, the configuration
gains volume energy because it is sampling a larger region in space
 with a lower energy density. The gain in volume energy grows linearly
(asymptotically) in one space dimension.

 The gradient terms, giving rise to a surface energy of
the droplet, saturate at about twice the kink mass in one space dimension.
To see this clearly, let us define a field configuration (droplet)
 parametrized by the radius $s$, as
\be
\Phi_{drop}(x-x_o, s) = \fisfa(x-x_o, s_o=s) \label{droplet}
\ee
and use $s$ as a parameter. The (classical) energy density as a function of
this
 parameter ${\cal{E}}(x,s)$ is depicted in figure 3 (a, b, c)
 for $s < s_o \; ;
s=s_o\; ; s>s_o$ respectively, for the case of the potential with a
 metastable minimum (figure 1(a)).

The total (classical) energy $E[s]$ is depicted in
figure (4)  for the potentials shown in figure (1).
The maximum of $E[s]$ for the case of a potential with a metastable
minimum (figure 5) is given by $s_o$ given by equation (\ref{so}),
 as this is the  value that corresponds
to the solution of the equation of motion for the static configuration,
that is, the ``sphaleron''.
The value $s_o$ corresponds to a ``critical droplet'', for $s<s_o$ the
droplet will shrink, as the cost in elastic surface term is greater than
the gain in volume energy. A droplet with $s=s_o$ is in unstable
equilibrium, whereas for $s>s_o$ the droplet will grow as the gain in
volume energy offsets the cost of elastic surface (wall) energy.

The sphaleron configuration corresponds to a {\it saddle}
point in functional space; there is one unstable direction corresponding
to the dilation or contraction of the droplet, one flat direction
corresponding to translational invariance, and the remaining (infinite)
directions are all stable, with the energy increasing quadratically for
small amplitudes away from the saddle.

This unstable direction in functional space plays a fundamental role:
consider an initial quantum mechanical state localized in the metastable
minimum, for example either a ground state wave functional or a thermal
density matrix for the quadratic potential centered on the metastable
minimum (see figure 1(a)). If the  probability for finding a droplet
with a size greater than the critical size in the initial state is
 non-zero,
the time evolution of this state will inexorably tend to spread the state
in such a way that field configurations which will sample the global
minimum in larger regions in space will acquire larger probabilities as
time evolves. This translates into the notion that the time evolved state
will give rise to a probability current at the saddle along the unstable
direction towards field configurations corresponding to growing droplets.
The ``decay rate'' of the initial state (or ensemble) will be determined
by the total probability current along this direction in functional
 space. Furthermore,
by translational invariance, these droplets appear with equal probability
at all points in space, and the total current passing through the
saddle point along the unstable direction will be proportional to the
volume.

At finite temperature, there will be a non-zero probability of finding
a critical (and larger) droplet in the initial ensemble and the
instability
will take the initial ensemble away from equilibrium.

 The ``decay''
of the initial ensemble will thus correspond to a process of thermal
activation, in which the initial state is driven ``over the barrier'' in
functional space. This barrier {\it should not be confused} with the
hump in the scalar potential, but it must be identified with the maximum
of $E[s]$ in figure 5 for the metastable case, that is to say, the
energy of a critical droplet.

We now study this process of thermal activation in {\it real time}
by following the time evolution of an initially prepared density matrix.

\section{\bf Real Time evolution of Initial Ensembles:}

We consider the situation in which there is a very rapid supercooling from
a high temperature phase in thermal equilibrium to a situation in which
the system gets trapped in a metastable state. This corresponds to the
case in which the initial ensemble is described by a density matrix in
thermal equilibrium, but ``centered'' at the metastable minimum of the
potential. We approximate this initial density matrix as that of a free
field theory centered at the metastable minimum and with particles with
mass $m$ determined by the second derivative of the potential at
 $\Phi_{-}$.
In the Schroedinger representation the density matrix elements are given
by
\ba
\rho(\Phi,\Phi',t=0)     & = & N \exp \left \{-\frac{1}{2}\int dx
 \int dy
 \left[(\Phi(x)-\Phi_{-})K_1(x-y)(\Phi(y)-\Phi_{-})+ \right. \right.
 \nonumber \\
                         &   &
 (\Phi'(x)-\Phi_{-})K_1(x-y)(\Phi'(y)-\Phi_{-}) \nonumber \\
                         &   & \left. \left.
 -2 (\Phi(x)-\Phi_{-})K_2(x-y)
 (\Phi'(y)-\Phi_{-})\right] \right\} \label{densitymatrix} \\ [0.1in]
 K_1(x-y)           & = & \int \frac{dk}{2\pi} e^{-ik(x-y)}
K_1(k) \;  \; ; \; \; K_1(k) =
\left[\frac{\omega_{k} \cosh[\beta\hbar\omega_{k}]}
{\hbar \sinh[\beta\hbar\omega_{k}]}\right] \label{kernel1} \\
K_2(x-y)            & = &  \int \frac{dk}{2\pi} e^{-ik(x-y)}
K_2(k) \; \; ; \; \; K_2(k) =
\left[\frac{\omega_{k}}
{\hbar \sinh[\beta\hbar\omega_{k}]}\right] \label{kernel2} \\
N                        & = & \left(Det\left[\frac{K_1-K_2}
{\pi}\right] \right)^{1/2} \; \; ; \; \;
\omega_{k}          =  \sqrt{k^2+m^2}  \label{norm}
\ea

We have chosen the normalization factor in such a way that
\be
Tr \rho = \int {\cal{D}}\Phi \rho(\Phi,\Phi) =1
\ee
With this normalization, the quantity $\rho(\Phi,\Phi)$ has the
interpretation of a probability density in functional space.

In terms of the Fourier components of the field $\Phi$, the above density
matrix
is easily identified with a density matrix in the coordinate
 representation
for a collection of independent harmonic oscillators with frequencies
$\omega_k$.

In particular, a meaningful question is: what is the probability (density)
to find the sphaleron configuration in the initial ensemble?.
 This quantity
is easily calculated in the high temperature limit where
 the Fourier transform
of $(\fisfa(x)-\Phi_{-})$ is non-negligible only for momenta
 $k \approx m$. Thus in the convolution of the sphaleron configuration
and the kernels, for temperatures $T \gg m$,
 the kernels in
the density matrix may be approximated by
\be
K_1(k)-K_2(k) = \frac{\omega_k}{\hbar}\tanh
\left[\frac{\hbar\omega_k}
{2k_BT}\right] \approx \frac{\omega_k^2}{2k_BT}
\ee

\noindent after some straightforward algebra we find in
 the ``thin-wall'' limit
\be
\rho(\fisfa,\fisfa) \simeq N \exp\left[-\frac{E_{sph}}{2k_BT}(1+12s_o)
\right]
\label{sfalprob}
\ee
with $s_o$ the radius of the droplet.

Then we see that the probability of finding the ``sphaleron''
configuration in the initial ensemble is {\it very different}
from $\exp[-E_{sph}/k_BT]$. The latter corresponds to a density
matrix of the form $\rho_0 = \exp[-(E_{sph}+H_q)/k_BT]$, with $H_q$
the Hamiltonian (\ref{hamsfal2}). This density matrix represents
 an ensemble
``centered'' at the sphaleron configuration.

We can also compute the ``average radius'' of the droplets in the
initial state. We find (see appendix B)
\ba
\langle \phi_{-1} \rangle & \approx & -\frac{3}{m} \sqrt{E_{sph}}
\label{averagefi} \\
\langle s \rangle         & \approx & s_o-\frac{3}{2} \label{averages}
\ea
in the thin wall limit $s_o \gg 1$ most of the ``droplets'' in the
initial ensemble are slightly smaller than the critical droplet.
This is clearly in agreement with the condition that our initial
ensemble is describing the metastable phase and the field configurations
that sample large regions of the stable phase are very rare.

The time evolution of the initial density matrix (\ref{densitymatrix})
is easily obtained in the quadratic approximation for (\ref{hamsfal2}),
that is setting $H_{I}=0$. The reason is that both the initial density
matrix and the Hamiltonian (around the sphaleron configuration) are
quadratic, thus at all times the density matrix will be of the gaussian
form. By expanding the field in the sphaleron basis and defining
\be
\Phi_{-}-\fisfa(x-x_o) =  \sum_{l}f_l(x-x_o) \bar{\phi}_l
 \; \; \; ; \; \; \;
\bar{\phi}_0           =  0 \label{phibarzero}
\ee
the initial density matrix becomes
\ba
\rho(\phi,\phi',t=0)  & = & N \exp \left\{-\frac{1}{2}
 \left[(K_1)_{l,l'}\left(\eta_{l}\eta_{l'}+
  {\eta'}_l {\eta'}_{l'}\right)
                       -2  (K_2)_{l,l'}\eta_l
 {\eta'}_{l'}\right] \right\} \label{densitymatrix2} \\
 (K_1)_{l,l'}    & = & \int dx \int dy
\int \frac{dk}{2\pi} e^{-ik(x-y)}K_1(k) f_l(x-x_o)f_{l'}(y-x_o)
 \label{kernel11} \\
(K_2)_{l,l'}     & = & \int dx \int dy
  \int \frac{dk}{2\pi} e^{-ik(x-y)}K_2(k) f_l(x-x_o)f_{l'}(y-x_o)
 \label{kernel22} \\
\eta_l                & = & (\phi_l-\bar{\phi}_l) \; \; ; \; \;
{\eta'}_l             =  ({\phi'}_l - \bar{\phi}_l)
\ea
Thus we see that in the sphaleron basis, the different modes are
mixed in the initial density matrix.

The time evolution of the density matrix is formally determined by
the Liouville equation
\be
i\hbar \frac{\partial \rho(\Phi,\Phi',t)}{\partial t} = \left[H,
\rho(\Phi,\Phi',t)\right] \label{liouville}
\ee
whose formal solution is
\be
\rho(\Phi,\Phi',t) = e^{-\frac{i}{\hbar}Ht}\rho(\Phi,\Phi',0)
e^{\frac{i}{\hbar}Ht} \label{timeevolution}
\ee

In the Schroedinger representation with $\Pi(x) =-i\hbar \delta /
\delta \Phi(x)$ it is straightforward to see that the functional
probability density obeys a continuity equation
\ba
\frac{\partial \rho(\Phi,\Phi,t)}{\partial t}
             &  =  & -\int dx \frac{\delta J[\Phi(x),t]}{\delta \Phi(x)}
 \label{continuity} \\
J[\Phi(x),t] &  =  &  \frac{-i\hbar}{2}\left(\frac{\delta}{\delta \Phi(x)}
-\frac{\delta}{\delta \Phi'(x)}\right)\rho(\Phi,\Phi',t)|_{\Phi=\Phi'}
\ea

In the ``sphaleron basis'',  the above continuity equation becomes
\ba
\frac{\partial \rho(\Phi,\Phi,t)}{\partial t}
             &  =  & -\sum_l \frac{\delta J_l[\phi,t]}{\delta \phi_l}
 \label{continuity2} \\
J_l[\phi,t]  &  =  &  \frac{-i\hbar}{2}\left(\frac{\delta}{\delta \phi_l}
-\frac{\delta}{\delta {\phi'}_l}\right)\rho(\phi,{\phi'},t)|_{\phi={\phi'}}
\label{probcurrent}
\ea
we have suppressed the label $(x_o)$, but we must keep in mind (and this
 will
become relevant later) that the $\phi_l$ depend on
it (though the $\bar{\phi}_l$ do not).

Following Langer\cite{langer}, we now identify the transition (nucleation)
 rate with the total probability current flowing across
the saddle point along the unstable direction (since the initial density
matrix has been normalized to one), i.e, the rate is given
 by
\be
\Gamma = \int^0_{-\infty}d\phi_{-1}\int_{-\infty}^{\infty}
{\prod}_{l \neq -1}d\phi_l \frac{\partial \rho(\phi,\phi,t)}{\partial t} =
-\int_{-\infty}^{\infty}J[\phi_{-1}=0;\phi_l]{\prod}_{l \neq -1}d\phi_l
\label{transitionrate}
\ee

The integral over the $l \neq -1$ modes (path integral) corresponds to
the {\it trace} of the time dependent density matrix in the reduced
functional space perpendicular to the unstable mode.

Although we can formally find the time evolution of the density
matrix by solving the Liouville equation with the Hamiltonian to
quadratic order in the sphaleron basis, the resulting equations of
motion mix the different modes and become very difficult to solve.

It becomes clear from the above expression that, in order to compute
 (\ref{transitionrate}),
we need to compute the ``reduced density matrix''
\be
\rho_r(\eta_{-1},{\eta'}_{-1},t) = {Tr}_{l \neq -1}\left(
e^{-\frac{i}{\hbar}H t}\rho(0)e^{\frac{i}{\hbar}Ht}\right)
\label{reduceddensitymatrix}
\ee
Because in the quadratic approximation around the sphaleron solution,
the Hamiltonian is a sum of  mutually commuting Hamiltonians for
each $l$, it is clear that the above trace and thus the reduced density
matrix evolves in time {\it only through} $H_{-1}$
\be
\rho_r(t) =  e^{-\frac{i}{\hbar}H_{-1} t}\rho_r(0)
e^{\frac{i}{\hbar}H_{-1}t}
\ee

 In terms of the
variable $\eta_l = \phi_l-\bar{\phi}_l$, the reduced density matrix
initially is
\be
\rho_r(\eta_{-1},{\eta'}_{-1},0) = {Tr'}\rho(0) =
\int_{-\infty}^{\infty} {\prod}_{l \neq
-1}d\eta_l\rho(\eta_{-1},\eta_l;\eta'_{-1},\eta_l)
\ee
The time evolution of the reduced density matrix will be solely determined
by the Liouville equation with the Hamiltonian corresponding to the
 unstable mode. Clearly, this is
a consequence of the quadratic approximation. Thus, by taking the
trace (integrating) over all functional directions perpendicular to the
unstable mode and obtaining a reduced density matrix for the unstable
coordinate, we have cast the multidimensional problem in terms of only
one quantum mechanical degree of freedom corresponding to the collective
coordinate representing the ``radius of the droplet''.
Now the problem becomes very similar to the quantum mechanical example
studied recently\cite{boyhol}.
It is at this stage that we recognize that by introducing a basis which
depends on the center of mass of the sphaleron $(x_o)$, we are not
treating translational invariance properly. For each $(x_o)$, the chosen
 basis
corresponds to an orthonormal coordinate system in functional space.
Changing $(x_o)$ amounts to performing a linear
 transformation on the coordinates $\phi^{[x_o]}_l$ (see appendix A),
 thus, by integrating over all the
possible values of the coordinates $\phi^{[x_o]}_l$ one recovers
translational invariance. However, one of the coordinates is only
integrated up to the saddle point and translational invariance is not
explicit in the above expression for the rate. To remedy this situation,
it is convenient at this point to perform a non-linear transformation
and treat $(x_o)$ as a collective coordinate. This is achieved by going
from the (cartesian) coordinate system $\{\phi_0 \; ,
 \phi_{l \neq 0}\}$ to  new
 (curvilinear) coordinates $\{x_o \; , \phi_{l \neq 0} \}$. There
is a Jacobian associated with this transformation. To leading
semiclassical order it is given by (see appendix A)
\be
{\cal{J}} = \sqrt{E_{sph}}+{\cal{O}}(\phi) \label{jacobian}
\ee
The collective coordinate is introduced into the trace (functional
integral) as\cite{gervaissakita,jasnowrudnick} (see appendix A)
\be
\int d\phi^{[x_o]}_0 \prod_{l \neq 0} d\phi_l = \int
d(x_o)\delta(\phi^{[x_o]}_0){\cal{J}} d\phi_0 {\prod}_{l \neq 0} d\phi_l
\label{collcoord}
\ee

 Postponing the integration over the collective
coordinate to the end of the calculation, the trace over the $l \neq -1$
modes can be
performed easily  and we find the reduced density matrix for the
unstable coordinate to leading
semiclassical order
\ba
\rho_r(\eta_{-1},\eta'_{-1}) & = & {\cal{N}}
 \exp\{-\frac{1}{2\hbar}[
\alpha(\eta_{-1}^2+{\eta'}_{-1}^2)+2\gamma \eta_{-1}{\eta'}_{-1}]\}
\label{reduceddensmat} \\
\frac{\alpha}{\hbar}         & = & (K_1)_{-1,-1} -\frac{1}{2}
\vec{Q}^T\tilde{K}^{-1}\vec{Q} \label{alfa} \\
\frac{\gamma}{\hbar}          & = & -(K_2)_{-1,-1}-\frac{1}{2}
\vec{Q}^T\tilde{K}^{-1}\vec{Q} \label{gamma} \\
Q_l                          & = & (K_1-K_2)_{-1,l} \; \; ; \; \;
l \neq 0, -1 \label{vectorQ} \\
\tilde{K}_{l,l'}             & = & (K_1-K_2)_{l,l'} \; \; ; \; \;
l,l' \neq 0, -1 \label{Ktilde} \\
{\cal{N}}                    & = & \sqrt{E_{sph}}
 \left(\frac{Det\left[\frac{K}{\pi}\right]}
{Det[\frac{\tilde{K}}{\pi}]}\right)^{\frac{1}{2}} \label{normalization}
\ea
with $K=(K_1-K_2)$ and $\tilde{K}$ is the same operator but without the row
and columns of matrix elements with the zero mode and the unstable mode
 in the ``sphaleron'' basis.
Since the reduced density matrix evolves in time only through $H_1$, the
continuity equation leads to (restoring the integral over the collective
coordinate)
\ba
& & \int dx_o {\cal{J}} \int_{-\infty}^{0}d\phi_{-1}
\frac{\partial \rho_r(\phi_{-1},\phi_{-1}, t)}{\partial t}
= - \int dx_o J_{-1}[\phi_{-1}=0,t]
\label{contequnstable} \\
& & J_{-1}[\phi_{-1},t]  =   \frac{-i\hbar{\cal{J}}}{2}\left(
\frac{\delta}{\delta \phi_{-1}}
-\frac{\delta}{\delta {\phi'}_{-1}}\right)
\rho_r(\phi_{-1},{\phi'}_{-1},t)|_{\phi_{-1}={\phi'}_{-1}}
\ea

The current at the saddle turns out to be translational invariant
(independent of $x_o$) because it only depends on $\bar{\phi}_{-1}$
(which is independent of $x_o$) and the integration over the collective
 coordinate will yield to a volume factor.

The remaining task is to find the time evolution of the reduced density
matrix (\ref{reduceddensmat}) with the Hamiltonian
(in the Schroedinger representation)
\be
 H_{-1} = \frac{1}{2}\left[-\hbar^2\frac{\delta^2}{\delta
\phi_{-1}^2} - \Omega^2 \phi^2_{-1}
\right] \nonumber
\ee

Since the initial reduced density matrix is quadratic and so is the
 evolution
Hamiltonian, we propose a gaussian ansatz for the time dependent reduced
density matrix (the label $\{-1\}$ is not explicitly written for
 $\eta$ to
avoid cluttering of notation)
\ba
\rho_r(\eta,\eta',t) & = & {\cal{N}}(t)
 \exp\left\{-\frac{1}{2\hbar}\left[
\alpha(t)\eta^2(t)+\alpha^{*}(t){\eta'}^2(t)+
2\gamma(t)\eta(t){\eta'}(t)\right]\right.+ \nonumber \\
                     &   &
\left. \frac{i}{\hbar}\left[P(t)\eta(t)-P^{*}(t){\eta'}(t)\right]\right\}
 \label{densmatanzatz} \\
             \eta(t) & = & \phi_{-1}-\bar{\phi}_{-1}(t) \; \; ; \; \;
             {\eta'}(t)  =  {\phi'}_{-1}-\bar{\phi}_{-1}(t)
             \label{timedepreddensmat}
\ea

Clearly, $\bar{\phi}_{-1}(t) \; ; \; P(t)$ are the expectation value of
 the
coordinate and canonical momentum along the unstable direction.
 The Liouville equation in the Schroedinger representation for the
$(-1)$
(unstable coordinate) now reads
\be
i\hbar \frac{\partial \rho_r(\eta,\eta',t)}{\partial t} =
\left[-\frac{\hbar^2}{2}\left(\frac{\delta^2}{\delta \phi_{-1}^2}-
\frac{\delta^2}{\delta {\phi'}_{-1}^2}\right) -\frac{1}{2}\Omega^2
\left(\phi_{-1}^2-{\phi'}_{-1}^2\right) \right]\rho_r(\eta,\eta',t)
\label{liouvilleevol}
\ee
The form of the above density matrix is dictated by hermiticity.
The time dependence of the parameters ${\cal{N}}(t), \phi_{-1}(t), P(t)$,
is obtained by comparing the left and right hand side of
 (\ref{liouvilleevol})
and matching the coefficients of the powers of $\eta \; ; \; {\eta'}$. We
find the following equations
\ba
i\frac{\dot{\cal{N}}}{{\cal{N}}} & = & \frac{1}{2} (\alpha-\alpha^*)
\label{normaleq} \\
-i\dot{\alpha}                   & = &
        -\alpha^2+\gamma^2-\Omega^2 \label{alfaeq} \\
-i\dot{\alpha}^*                 & = &
{\alpha^*}^2-\gamma^2+\Omega^2 \label{alfastareq} \\
i\dot{\gamma}                    & = &  \gamma(\alpha-\alpha^*)
\label{gammaeq} \\
\dot{P}                          & = & \Omega^2\bar{\phi}_{-1}
\; \; ; \; \; P=P^* \label{momentumeq} \\
\dot{\bar{\phi}}_{-1}                  & = & P \label{classieq}
\ea
The last two equations are the classical equations of motion for
 the unstable
coordinate in an inverted harmonic oscillator.
The boundary conditions are:

\[ P(0) = 0 \; ; \; \bar{\phi}_{-1}(0)=\bar{\phi}_{-1} \; \; \;
 {\cal{N}}(0)
= {\cal{N}} \; ; \; \alpha(0)=\alpha^*(0)=\alpha \; ; \;
 \gamma(0)=\gamma \]

In terms of the real and imaginary parts of $\alpha$ we find the solution
to the above equations to be
\ba
\alpha_I(t)   & = & -\frac{\omeg\sinh[2\omeg t]}
{\cosh[2\omeg t]-\cos(2\delta)}
\label{alfaim} \\
\alpha_R(t)   & = & \alpha(0)\left[\frac{1-\cos(2\delta)}{\cosh[2\Omega t]-
\cos(2\delta)}\right] \label{alfareal}\\
\gamma(t)     & = & \gamma(0)\left[\frac{1-\cos(2\delta)}{\cosh[2\Omega t]-
                   \cos(2\delta)}\right] \label{gammat}     \\
{\cal{N}}(t)  & = & {\cal{N}}(0)\left[\frac{1-\cos(2\delta)}{\cosh[2\Omega t]-
                   \cos(2\delta)}\right]^{\frac{1}{2}} \label{normt} \\
\bar{\phi}_{-1}(t) & = & \bar{\phi}_{-1}(0)\cosh[\Omega t] \; \;
 ; \; P(t) =
             \dot{\bar{\phi}}_{-1}(t) \label{classolu} \\
\tan\delta    & = & \frac{\Omega}{\sqrt{\alpha_R^2(0)-\gamma^2(0)}}
= \frac{\Omega}{W}
\label{tandelta}
\ea
We obtain the expression for the rate per unit volume
\ba
& &\frac{\Gamma(t)}{L} = J[\bar{\phi}_{-1}, t] \\
& &\frac{\Gamma(t)}{L} = -\Omega \bar{\phi}_{-1}(0)A(t)
 \sqrt{E_{esp}}{\cal{N}}(0)
  \exp \left\{-\frac{1}{\hbar}\bar{\phi}_{-1}^2(0)(\alpha_R(0)+\gamma(0))
B(t) \right\} \label{finalrate} \\
& & A(t) = \frac{W^2}{\Omega^2} \frac{\sinh[\Omega t]}{\left[\cosh^2[
\Omega t] + \frac{W^2}{\Omega^2}\sinh^2[\Omega t]\right]^{\frac{3}{2}}}
 \label{prefaoft} \\
& & B(t) =  \frac{1}{\left[1+\frac{W^2}{\Omega^2}
\tanh^2[\Omega t]\right]} \label{boft}
 \ea

There are two competing effects that lead to the final expression for
the prefactor $A(t)$ in the rate. The first corresponds to the
``rolling'' of the expectation value of the unstable coordinate
($\bar{\phi}_{-1}$)
 down the inverted quadratic potential,
the canonical momentum of this mode contributes to the prefactor a term
proportional to $\sinh[\Omega t]$. The second contribution has its
origin in the spread of the probability distribution function (width
of the gaussian) associated with the growth of the unstable fluctuations
contributing typical factors $\sinh[2\Omega t]$.
The width of the reduced gaussian density matrix gives the two-point
correlation function of the fluctuation of the unstable mode. This
correlation function grows exponentially because of the instability.
 Thus we see that the
spread of the probability distribution is the dominant term in the
time dependent rate. This observation is in agreement with the arguments
of classical homogeneous nucleation in that  the fluctuations are the
main responsibles for the growth of droplets and the decay of the
metastable state.

 In order to find a more compact expression for the rate we need to
obtain  the
 coefficients $\alpha(0)\; ; \; \gamma(0)\; ; \; \bar{\phi}_{-1}(0)$.
 In appendix B we find a simple
 expression for these coefficients and compute them in the high
 temperature  limit and in the ``thin-wall'' approximation. Finally
 our expression for the rate is
 \ba
 & &\frac{\Gamma(t)}{L} =  3\omeg \left(\frac{E_{sph}k_BT}{\pi m
\hbar^2}\right)
 {\cal{D}} A(t)  \exp\left\{-\frac{CE_{sph}}{k_BT}
B(t) \right\} \label{truerate}
\ea
 with $C \approx 5.1759$ (see appendix B) and
 where we have absorbed temperature and ($\hbar$)
 factors in the kernels (making them dimensionless) and defined
 \ba
 {\cal{D}} & = &  \left[\frac{Det(D)}{Det(D'')}\right]^{\frac{1}{2}}
 \nonumber \\
 D         & = & \prod_k \left[\beta \hbar \omega_k\right]
 \tanh\left[\frac{\beta \hbar \omega_k}{2}\right] \nonumber
\ea
and $D''$ the same operator in the sphaleron basis but without the rows and
columns corresponding to the zero mode and the unstable mode.
The rate as a function of time is depicted in figure (6) for an arbitrary
choice of the parameters. We see that the rate vanishes at $t=0$, reaches
a maximum and eventually falls off exponentially to zero at very long
times.
In the thin wall limit, the maximum occurs at $t \approx 1/W \approx
1/ m $.
 The reason why the rate vanishes at $t=0$ is because the initial
state needs to spread out to reach the saddle and that at the initial
time the expectation value of the canonical momentum conjugate to
the unstable coordinate vanishes and that
 the density matrix has real kernels. Thus the current vanishes at $t=0$.

The fact that at very large
times the rate must fall off to zero is a consequence of unitary time
evolution. The total probability is conserved.  As current is passing
over the saddle, the probability density in the metastable state is
depleted and thus the current must diminish.

This is the main result of this work, a time dependent rate that
incorporates the realistic condition of a supercooled non-equilibrium
 initial state. This result is certainly very different from the familiar
approach to decay of metastable states described as steady state processes
in thermal equilibrium.  We want to emphasize the main
differences with previous work on this problem: our expression for the
rate obtains from a {\it real time evolution} of an initial supercooled
state (ensemble).
 The vanishing of the rate at early times is a consequence of the
zero average canonical momentum and real kernels of the initial ensemble
(compatible with local thermodynamic equilibrium), and the
vanishing of the rate at late times is an unavoidable consequence of
unitary time evolution and the depletion of probability in the initial
state.

 Certainly, our approximation
of keeping the quadratic fluctuations around the sphaleron configuration
is not valid at long times as the amplitudes of the fluctuations become
very large and the form of the decay rate obtained will not be accurate
at long times. However,  unitary time evolution and conservation of
probability will necessarily
constrain the {\it time dependent rate} to fall-off to zero at long
times.

This clearly illuminates the fact that the familiar results for the
decay rate correspond to the very particular situation of a steady
state and replenishing of probability in the metastable state.

In a remarkable experiment on nucleation of classical fluids under
shear, Min and Goldburg have recently reported\cite{mingold}
 a {\it time
dependent nucleation rate} that is strongly dependent on the initial
conditions (shear). The nucleation rate reported by these authors
has a qualitative behavior very similar to
the time dependent rate (\ref{truerate}), vanishing at early
times, reaching a maximum and falling off at late times. Although
our calculation is clearly quantum mechanical, we conjecture that its
classical limit will offer a description of classical statistical
 systems.

\section{Particle Production:}

We are now in condition to understand the production of particles of
``true vacuum'' as the metastable phase decays into the true phase
via the process of thermal activation. Consider the  operators that
create and annihilate particles of momentum $\vec{k}$
 of the ``true vacuum'', that is the
 vacuum centered at $\Phi_+$. Quantizing in a volume $L$
these are given by
\ba
a^{\dagger}_k  & = & \frac{1}{\sqrt{2\hbar L}}\int dx e^{ikx}
\left[\sqrt{\omega^+_k}(\Phi(x)-\Phi_+) -\frac{i}{\sqrt{\omega^+_k}}
\Pi(x)\right] \label{creationop} \\
a_k            & = & \frac{1}{\sqrt{2\hbar L}}\int dx e^{-ikx}
\left[\sqrt{\omega^+_k}(\Phi(x)-\Phi_+) +\frac{i}{\sqrt{\omega^+_k}}
\Pi(x)\right] \label{destrucop}
\ea
with $\omega^+_k$ the frequencies of harmonic oscillator quanta around
the ``true vacuum''.
Now we can expand the field and its canonical momentum in the
``sphaleron'' basis by writing
\[ \Phi(x)-\Phi_+ = \Phi(x)-\Phi_{-}+\Delta\Phi \]
Since we expect that the maximum contribution to the rate of particle
production will arise from the  evolution of the unstable mode, we
will only keep this contribution to the creation and annihilation
 operators. The average number of ``true'' particles as a function of
time is then given by
\be
\langle N_k \rangle (t) =
 \frac{Tr a^{\dagger}_k a_k \rho(t)}{Tr\rho(t)} =
 \frac{Tr a^{\dagger}_k (t) a_k (t) \rho(0)}{Tr\rho(0)}
 \ee
where $a^{\dagger}_k (t) \; ; \;  a_k (t)$ are the operators in the
Heisenberg picture with the Hamiltonian (\ref{hamsfal2}).

Using, in this approximation
\ba
\phi_{-1}(t) & = & \phi_{-1}(0) \cosh[\omeg t] +
 \frac{\pi_{-1}(0)}{\omeg} \sinh[\omeg t]
  \nonumber \\
\pi_{-1}(t)  & = & \dot{\phi}_{-1}(t)
 \nonumber
\ea
and that in the initial density matrix
\ba
 \langle \eta_{-1} \rangle(0) & = & \langle \pi_{-1}(0) \rangle =0
\nonumber \\
\langle \eta^2_{-1} \rangle(0)& = & \frac{\hbar}{2(\alpha(0)+\gamma(0))}
\nonumber \\
 \langle \pi^2_{-1}(0) \rangle& = & \frac{\hbar}{2(\alpha(0)-\gamma(0))}
\nonumber
\ea
and using the results obtained in the appendices we find the leading
contribution of the unstable evolution to the total number of particles
of true vacuum produced as a function of time (we are neglecting time
independent and subleading contributions)
\ba
 \sum_k \langle N_k(t) \rangle \approx
\frac{3\pi^2}{4}\left(\frac{k_BT}{\hbar m}\right) & &
\left\{ \cosh^2[\omeg t]
\left[S_1\left(\frac{m^+}{m}\right)+
S_2\left(\frac{m^+}{m}\right)\left(\frac{9E_{sph}}{k_BT}\right)\right]
\right. \nonumber \\
                                                  & & \left.
                                                  + \sinh^2[\omeg t]
 S_3\left(\frac{m^+}{m}; \frac{\omeg}{m^+}\right)\right\}
\ea

The functions $S_1 \; ; \; S_2 \; ; \; S_3$ are determined by the
 structure
factor of the unstable mode $f_{-1}(x-x_o)$, the unstable frequency
$\omeg$
 and the frequencies of the quanta in the true vacuum.

In the thin wall limit the momentum integrals that determine these
structure factors are all dominated by a peak at $k \approx 0$ of
width $\approx m$. In this limit $m^+ / m^- \approx 1 \; ; \; \Omega /
m \ll 1$, in this limit these three functions become pure numbers of
${\cal{O}}(1)$.

\vspace{3mm}

{\bf Validity of the Approximations:}

Our result for the transition rate for thermal activation relies on
several approximations.

 The first one corresponds to keeping the
quadratic  fluctuations around the droplet (``sphaleron'') configuration.
This approximation provides the leading semiclassical expression for
the rate and is justified at early and intermediate times, since the
fluctuations of the unstable coordinate will grow typically as
$\cosh[2\Omega t]$ at early times the amplitud for the fluctuation will
grow and presumably higher order terms (cubic and quartic) may have to
be kept to fully understand the long time behavior.

The thin wall approximation is justified in the case of a small
supercooling (small energy density difference between the metastable
and the stable state), this condition may be relaxed in the case of
strong supercooling. In this latter case, the ``droplet'' configuration
will be indistinguishable from a localized large amplitude fluctuation.

The initial state condition will be justified in the case of a very
rapid supercooling. Since the maximum of the rate occurs at a time
$t \approx 1 / m$ (in the thin wall limit) the initial supercooled
state condition will be justified if the time during which the
supercooling takes place is much smaller than $1/m$.
 Clearly this situation is not general
and will have to be understood case by case. In particular in the
case of inflationary cosmologies, this inequality results in that the
Hubble constant must be much larger than the mass of the quanta in the
metastable state. This seems to be the case in the most popular theories
on inflation that require a  supercooled first order phase transition.

Clearly at low temperatures (much smaller than $E_{sph}$) tunneling
will be the most important mechanism for metastable decay. Eventually
there will be a crossover between tunneling and thermal activation that
must be understood better. We must say however that within the gaussian
approximation for the fluctuations around the sphaleron, our solution
for the time evolution of the density matrix {\it does incorporate
quantum corrections} as may be seen from our result prior to taking the
high temperature (classical) limit.

Although we presented the analysis in a $1+1$ dimensional field theory,
we see no problem in considering our approach in $3+1$ dimensions. We
are currently extending our results to the three dimensional case and
expect to report on it in a forthcoming article.

\newpage

\section{\bf Conclusions and Implications}

In this article we have studied the process of decay of a metastable
state via thermal activation mediated by ``sphaleron'' (droplet)
configurations. We offered a {\it real time analysis} based on the
time evolution of an initially prepared density matrix corresponding to
a supercooled state. The initial density matrix is assumed to describe
the harmonic oscillations of the metastable state at an initial temperature
much larger than the mass of the quanta in this state but smaller
than the energy of the sphaleron configuration.

We obtain a time dependent nucleation rate. This rate vanishes at early
times as a consequence of the quantum mechanical spreading of the density
matrix, reaches a maximum at a time $t \approx 1/m$ with $m$ the mass
of the quanta in the metastable state and vanishes at long times as a
consequence of unitary time evolution. The most important contribution
to the rate arises from the fluctuations along the unstable direction in
functional space.

This behavior is similar to nucleation rates obtained experimentally
recently in classical fluids under shear, but strikingly different from
the familiar expressions for the decay rate. The difference with the
usual result is that it corresponds to a {\it steady state} process in
thermal equilibrium in which the metastable state is replenished at the
same rate at which probability is flowing across the saddle.

We also provided an estimate for the number of particles of the stable
state produced as a function of time during the process of decay of the
metastable state.

For applications in the early universe or to describe processes that
involve baryon number violation one must understand better the physics of
the initial conditions. We believe that a steady state assumption, leading
to the familiar result for the rate is not warranted in either case and
considering an initially supercooled state may be closer to the physical
situation.

Within the context of the inflationary scenario, our results offer a
rather pessimistic outlook. It is widely accepted that in
 order to complete the phase transition a fairly large nucleation rate is
required at long times, precisely when unitarity forces the
rate to vanish. The actual time at which the rate starts to fall off as
as a consequence of unitarity will depend on the details of the potentials,
and the amount of supercooling. A deeper understanding of this time regime
will require to go beyond the quadratic approximation.

With respect to sphaleron mediated baryon number violating processes,
a time dependent rate may have important consequences for obtaining
a net baryon asymmetry.
 In particular, when the thermal activation
rate falls below the expansion rate in an expanding cosmology, sphaleron
transitions go out of equilibrium. If furthermore the CP violating
 effects may be accomodated in the theory, the necessary elements for
baryon number violation are in place,
 but the question must be addressed within the
realm of a realistic gauge theory with fermions. The influence of
fermions on the dynamics of the droplets has not received much attention
but will perhaps be an important ingredient for a deeper understanding
of these processes.

\vspace{5mm}

{\bf Acknowledgments}
The authors would like to thank
R. Holman, D. Jasnow, W. Goldburg, J. Kapusta and R. Willey
 for very illuminating conversations.
D.B. would like to thank A. Weldon and  H. de Vega for
 discussions and suggestions. The authors acknowledge support through a
binational collaboration supported by N.S.F. and CNPq, through Grant
No: INT-9016254 at the University of Pittsburgh. D.B. also acknowledges
support from N.S.F. through Grant No: PHY-8921311. C. Arag\~{a}o de
Carvalho thanks the Physics Dept. at the Univ. of Pittsburgh,
 and D.B. thanks the Departamento  de Fisica Universidade Pontificia
Catolica for hospitality.

\section{\bf Appendix A:}
Here we collect some relevant results for the treatment of collective
coordinates. From the expansion around the sphaleron solution centered
at $x_o$ (\ref{fil}) one  finds
\ba
\phi^{[x_o]}_l & = & \int dx f_{l}(x-x_o)
\left(\Phi(x)-\Phi_{sph}(x-x_o)\right) \label{phixol} \\
\frac{d \phi^{[x_o]}_l}{dx_o}
              & = & \int dx f_{l}(x-x_o)\frac{d \Phi(x)}{dx}=
\sqrt{E_{sph}}\delta_{l,0}+\sum_{l'}\int dx f_{l}(x-x_o)
\frac{df_{l'}(x-x_o)}{dx}\phi^{[x_o]}_{l'} \label{derphil}
\ea
where we have used (\ref{zeromode}). Treating $x_o$ as a collective
coordinate
implies\cite{jackiw,christ,tomboulis,creutz,callan,parmentola,rajaraman}
 the expansion
\be
 \Phi(x) = \fisfa(x-x_o)+ \sum_{l \neq 0}
 \phi_{l}f_{l}(x-x_o) \label{colcorexp}
 \ee
 and treating $\{x_o \; ; \; \phi_{l \neq 0}\}$ as coordinates.
In the functional integrals (traces) the change over to collective
coordinates is done by introducing\cite{gervaissakita,jasnowrudnick}
\ba
1 & = & \int dx_o \delta\left(\phi^{[x_o]}_0\right){\cal{J}}
 \label{fadepop} \\
{\cal{J}}
  & = & \left\|\frac{d\phi^{[x_o]}_0}{dx_o}\right\|
 \approx \sqrt{E_{sph}}+{\cal{O}}
(\phi)   \label{jacob}
\ea

\section{\bf Appendix B:}

In this appendix we derive an expression for the coefficients
 $\alpha(0) \; ; \; \gamma(0)$ that enter in the reduced density matrix.
{}From the relations (\ref{alfa},\ref{gamma}) we find
\ba
\frac{1}{\hbar}(\alpha(0)-\gamma(0)) & = &
 \int \frac{dk}{2\pi\hbar}
\int dx \int dy f_{-1}(x-x_o)f_{-1}(y-x_o)e^{-ik(x-y)} \times
\nonumber \\
                                     &   &  \omega_k
\left(\frac{\cosh[\beta\hbar \omega_k]+1}
{\sinh[\beta\hbar \omega_k]}\right)
\ea
since the Fourier transform of $f_{-1}(x)$ is localized in k-space, we
can use the high temperature expansion of the kernels to find in the
high temperature limit $\beta m \ll 1$
\be
\frac{1}{\hbar}(\alpha(0)-\gamma(0)) \approx \frac{2k_BT}{\hbar^2}
\label{alfaminusbeta}
\ee

We also need the sum of the coefficients; it is obtained as follows:
for a {\it fixed} $x_o$ consider the quantity
 (here $\eta = \eta^{[x_o]}_{-1}$)
\ba
& &\int dy f_{-1}(x-x_o)f_{-1}(y-x_o) \langle (\Phi(x)-\Phi_{-})
(\Phi(y)-\Phi_{-}) \rangle
 = \langle \left(\eta(0)^2\right) \rangle \nonumber \\
& &\langle \left(\eta^{[x_o]}(0)\right)^2 \rangle
          =  \frac{\int {\prod}_l d\phi^{[x_o]}_l (\eta^{[x_o]})^2
         \rho(\phi^{[x_o]}_l,\phi^{[x_o]}_l)}
         {\int {\prod}_l d\phi^(x_o)_l
         \rho(\phi^{[x_o]}_l,\phi^{[x_o]}_l)}
\label{expect}
\ea
Where the expectation value is in the initial density matrix.
 Introducing
the collective coordinate as in (\ref{collcoord},\ref{jacobian}),
 and performing
the integrals over all directions perpendicular to $l=-1$,
 this expectation
value can be recast in leading semiclassical order as
\be
\frac{\int dx_o \int d\phi^{[x_o]}_{-1}(\eta^{[x_o]}_{-1})^2
\rho_r(\phi_{-1},\phi_{-1})}{\int dx_o \int d\phi^{[x_o]}_{-1}
\rho_r(\phi_{-1},\phi_{-1})} = \frac{\hbar}{2(\alpha(0)+\gamma(0))}
\ee
Thus
\be
\frac{\hbar}{2(\alpha(0)+\gamma(0))}= \int dx \int dy \int \frac{dk}{2\pi}
f_{-1}(x-x_o)f_{-1}(y-x_o)\frac{\hbar e^{-ik(x-y)}}{2\omega_k \tanh
[\beta \hbar \omega_k/2]}
\ee
Again, the Fourier transform may be calculated in the high temperature limit
because the mode functions are localized in k-space; using
$\tanh[\beta\hbar\omega_k/2] \approx \beta\hbar\omega_k/2$,
 we finally find
\be
\frac{1}{\hbar}(\alpha(0)+\beta(0)) \approx \frac{m^2}{9k_BT}
\left(\frac{3\pi}{2h(s_o)}\right)
\ee
With $h(s_o)$ a function of only the radius of the droplet.
 In the thin-wall
approximation, we find numerically

\[C= \left(\frac{3\pi}{2h(s_o \gg 1)}\right)  \approx 5.1759 \]

We also find in the thin-wall approximation that
\ba
\bar{\phi}_{-1}(0) & \approx &  -\frac{3}{m}\sqrt{E_{sph}}
\label{phi0} \\
\tan\delta         & \approx & \delta \approx \frac{3}{\sqrt{2C}}
\frac{\Omega}{m} \label{delta}
\ea

\newpage

\underline{\bf Figure Captions:}
\vspace{2mm}

\underline{\bf Figure 1}:
$V(\Phi)$ vs. $\Phi$ for when $\Phi_{-}$ corresponds to:
  a metastable state (a), a degenerate state (b), and the true vacuum
state (c).

\vspace{2mm}

\underline{\bf Figure 2}:
$\Phi_{sph}(x)$ vs $x$ for a ``thin-wall'' sphaleron (droplet).

\vspace{2mm}

\underline{\bf Figure 3}:

Energy density as a function of the radius of the sphaleron
${\cal{E}}(x;s)$ in the metastable case for $s < s_o$ (a); $s=s_o$ (b)
and $s > s_o$ (c).

\vspace{2mm}

\underline{\bf Figure 4}:

Total energy of a droplet $E(s)$ as a function of  the ``radius'' $s$,
for the cases in which $\Phi_{-}$ represents
the metastable state (a), degenerate state (b) and true vacuum state
(c). (See figure 1).

\vspace{2mm}

\underline{\bf Figure 5}:

Total energy of a droplet configuration $E(s)$ as a function of $s$
 detailing the maximum at $s= s_o$.

\vspace{2mm}

\underline{\bf Figure 6}:

Decay rate per unit volume $\Gamma(t)/L$ as a function of $Wt$ for the
arbitrary values $C E_{sph}/k_{B}T = 2.0$, $\Omega / W = 0.01$. (The
constant coefficients in the expression for the rate were set to 1).

\end{document}